\documentstyle[11pt]{article}
\begin{document}
%\tighten
\textheight 9.4in
\textwidth 6.0in
\topmargin -0.2truecm
\title{ \hfill
%hep-ph/0203196\\ 
\vskip 0.01truecm {\large{\bf Constraining R violation from Anomalous
Abelian Family symmetry} \footnote{Talk given at 7th Workshop on High
Energy Particle Phenomenology, HRI, Allahabad, India, and  
XIV DAE Symposium on High Energy Physics, hosted by 
University of Hyderabad, India. To appear in the proceedings of the
Symposium. The work was done in collaboration with Anjan
Joshipura and Sudhir Vempati.}}}
\vskip 0.1truecm

\author{ Rishikesh Vaidya  \\
{\ns\it Theoretical Physics Group, Physical Research Laboratory,}\\
{\ns\it Navarangpura, Ahmedabad, 380 009, India.}}
\date{}
%------------------------------------------------------------
\def\ap#1#2#3{           {\it Ann. Phys. (NY) }{\bf #1} (19#2) #3}
\def\arnps#1#2#3{        {\it Ann. Rev. Nucl. Part. Sci. }{\bf #1} (19#2) #3}
\def\cnpp#1#2#3{        {\it Comm. Nucl. Part. Phys. }{\bf #1} (19#2) #3}
\def\apj#1#2#3{          {\it Astrophys. J. }{\bf #1} (19#2) #3}
\def\asr#1#2#3{          {\it Astrophys. Space Rev. }{\bf #1} (19#2) #3}
\def\ass#1#2#3{          {\it Astrophys. Space Sci. }{\bf #1} (19#2) #3}

\def\apjl#1#2#3{         {\it Astrophys. J. Lett. }{\bf #1} (19#2) #3}
\def\ass#1#2#3{          {\it Astrophys. Space Sci. }{\bf #1} (19#2) #3}
\def\jel#1#2#3{         {\it Journal Europhys. Lett. }{\bf #1} (19#2) #3}

\def\ib#1#2#3{           {\it ibid. }{\bf #1} (19#2) #3}
\def\nat#1#2#3{          {\it Nature }{\bf #1} (19#2) #3}
\def\nps#1#2#3{          {\it Nucl. Phys. B (Proc. Suppl.) }
                         {\bf #1} (19#2) #3}
\def\np#1#2#3{           {\it Nucl. Phys. }{\bf #1} (19#2) #3}
\def\pl#1#2#3{           {\it Phys. Lett. }{\bf #1} (19#2) #3}
\def\pr#1#2#3{           {\it Phys. Rev. }{\bf #1} (19#2) #3}
\def\prd#1#2#3{           {\it Phys. Rev. }{\bf #1} (20#2) #3}
\def\prep#1#2#3{         {\it Phys. Rep. }{\bf #1} (19#2) #3}
\def\prl#1#2#3{          {\it Phys. Rev. Lett. }{\bf #1} (19#2) #3}
\def\pw#1#2#3{          {\it Particle World }{\bf #1} (19#2) #3}
\def\ptp#1#2#3{          {\it Prog. Theor. Phys. }{\bf #1} (19#2) #3}
\def\jppnp#1#2#3{         {\it J. Prog. Part. Nucl. Phys. }{\bf #1} (19#2) #3}

\def\rpp#1#2#3{         {\it Rep. on Prog. in Phys. }{\bf #1} (19#2) #3}
\def\ptps#1#2#3{         {\it Prog. Theor. Phys. Suppl. }{\bf #1} (19#2) #3}
\def\rmp#1#2#3{          {\it Rev. Mod. Phys. }{\bf #1} (19#2) #3}
\def\zp#1#2#3{           {\it Zeit. fur Physik }{\bf #1} (19#2) #3}
\def\fp#1#2#3{           {\it Fortschr. Phys. }{\bf #1} (19#2) #3}
\def\Zp#1#2#3{           {\it Z. Physik }{\bf #1} (19#2) #3}
\def\Sci#1#2#3{          {\it Science }{\bf #1} (19#2) #3}
\def\n.c.#1#2#3{         {\it Nuovo Cim. }{\bf #1} (19#2) #3}
\def\r.n.c.#1#2#3{       {\it Riv. del Nuovo Cim. }{\bf #1} (19#2) #3}
\def\sjnp#1#2#3{         {\it Sov. J. Nucl. Phys. }{\bf #1} (19#2) #3}
\def\yf#1#2#3{           {\it Yad. Fiz. }{\bf #1} (19#2) #3}
\def\zetf#1#2#3{         {\it Z. Eksp. Teor. Fiz. }{\bf #1} (19#2) #3}
\def\zetfpr#1#2#3{         {\it Z. Eksp. Teor. Fiz. Pisma. Red. }{\bf #1} (19#2) #3}
\def\jetp#1#2#3{         {\it JETP }{\bf #1} (19#2) #3}
\def\mpl#1#2#3{          {\it Mod. Phys. Lett. }{\bf #1} (19#2) #3}
\def\ufn#1#2#3{          {\it Usp. Fiz. Naut. }{\bf #1} (19#2) #3}
\def\sp#1#2#3{           {\it Sov. Phys.-Usp.}{\bf #1} (19#2) #3}
\def\ppnp#1#2#3{           {\it Prog. Part. Nucl. Phys. }{\bf #1} (19#2) #3}
\def\cnpp#1#2#3{           {\it Comm. Nucl. Part. Phys. }{\bf #1} (19#2) #3}
\def\ijmp#1#2#3{           {\it Int. J. Mod. Phys. }{\bf #1} (19#2) #3}
\def\ic#1#2#3{           {\it Investigaci\'on y Ciencia }{\bf #1} (19#2) #3}
\def\tp{these proceedings}
\def\pc{private communication}
\def\ip{in preparation}
\relax
\newcommand{\GeV}{\,{\rm GeV}}
\newcommand{\MeV}{\,{\rm MeV}}
\newcommand{\keV}{\,{\rm keV}}
\newcommand{\eV}{\,{\rm eV}}
\newcommand{\Tr}{{\rm Tr}\!}
\renewcommand{\arraystretch}{1.2}
\newcommand{\beq}{\begin{equation}}
\newcommand{\eeq}{\end{equation}}
\newcommand{\beqa}{\begin{eqnarray}}
\newcommand{\eeqa}{\end{eqnarray}}
\newcommand{\ba}{\begin{array}}
\newcommand{\ea}{\end{array}}
\newcommand{\bmat}{\left(\ba}
\newcommand{\emat}{\ea\right)}
\newcommand{\refs}[1]{(\ref{#1})}
\newcommand{\ler}{\stackrel{\scriptstyle <}{\scriptstyle\sim}}
\newcommand{\ger}{\stackrel{\scriptstyle >}{\scriptstyle\sim}}
\newcommand{\lag}{\langle}
\newcommand{\rag}{\rangle}
\newcommand{\ns}{\normalsize}
\newcommand{\cm}{{\cal M}}
\newcommand{\gr}{m_{3/2}}
\newcommand{\p}{\partial}
\def\u1x{ $U(1)_{X}$}
\def\rp{ $R_P$}
\def\321{$SU(3)\times SU(2)\times U(1)$}
\def\tl{{\tilde{l}}}
\def\tL{{\tilde{L}}}
\def\bd{{\overline{d}}}
\def\tL{{\tilde{L}}}
\def\a{\alpha}
\def\b{\beta}
\def\g{\gamma}
\def\c{\chi}
\def\d{\delta}
\def\D{\Delta}
\def\db{{\overline{\delta}}}
\def\Db{{\overline{\Delta}}}
\def\e{\epsilon}
\def\l{\lambda}
\def\n{\nu}
\def\m{\mu}
\def\nt{{\tilde{\nu}}}
\def\p{\phi}
\def\P{\Phi}
\def\x{\xi}
\def\r{\rho}
\def\s{\sigma}
\def\t{\tau}
\def\th{\theta}
\def\ne{\nu_e}
\def\nm{\nu_{\mu}}
\def\rp{$R_P$}
\def\mp{$M_P$}
\def\emt{$L_e-L_{\mu}-L_{\tau}$}
\renewcommand{\Huge}{\Large}
\renewcommand{\LARGE}{\Large}
\renewcommand{\Large}{\large}
\maketitle
%\vskip 2.0truecm
\begin{center}
%\underline{\bf{ABSTRACT}}
\end{center}

%\vskip 0.4truecm
\begin{abstract}
The patterns of $R$ violation resulting from imposition of a gauged
$U(1)$ horizontal symmetry,  on the minimal supersymmetric standard model
are
systematically analyzed. We concentrate on a  class of models
with integer $U(1)$ charges chosen to reproduce the quark masses and
mixings as
well as charged lepton masses exactly or approximately.
The $U(1)$ charges are further restricted from the requirement that
very large bilinear lepton number violating terms should not be allowed
in the super-potential. It is shown that 
all 
the trilinear $\lambda'_{ijk}$ and all but at most two trilinear
$\lambda_{ijk}$ couplings vanish or are enormously suppressed. 
\end{abstract} 

\section{Introduction}
        Supersymmetric Standard Models without R-parity contain
        large number (48) of free R-violating parameters and hence
        lack predictive power. However, models without R-parity
        are interesting in their own right in many ways. Models
        without lepton number can naturally accommodate neutrino masses
        as required by the present solar and atmospheric neutrino
        anomalies \cite{rneutrino}. In such cases,
the structure of lepton number       
        violating couplings plays an important role as it determines
        the pattern of the mixing between the neutrino states.

        Since these trilinear L violating couplings are similar to the standard Yukawa couplings,
        it is generally argued that they are also
        hierarchical in nature. Given that the hierarchical nature of
        the standard Yukawa couplings can be attributed to an unbroken
        symmetry at a high scale, it would be interesting to study the
        structure of R-violating couplings under the influence of such
        a symmetry. In this talk, we present a systematic study
        of R-violating interactions in the framework of a U(1) family
        symmetry and their allowed patterns from low energy phenomenology.
\section{U(1) SYMMETRY AND THE $\e$ PROBLEM}
The ratio of masses of fermions belonging to different 
generations (in quark and lepton sector both) when expressed in 
terms of Cabibo angle $\l$ =0.22, 
show a geometrical hierarchy. A mechanism that sets this
orders of magnitude could be a broken U(1) family symmetry originally
suggested by Froggatt and Nielsen~(FN) \cite{FN}.
We exploit the  FN mechanism to shed light on the structures of R
violating couplings \cite{rishi}.

Let us consider the MSSM augmented with a gauged horizontal  U(1) symmetry.
The standard super-fields $(L_{i},Q_{i},D_{i},U_{i},E_{i},H_{1},H_{2})$
are assumed to carry the charges $(l_{i},q_{i},d_{i},u_{i},e_{i},h_{1},h_{2})$
respectively with i running from 1 to 3. The  U(1) symmetry is assumed to be broken at a high scale by
the vacuum expectation value (vev) of one gauge singlet super-field $\theta$
with the U(1) charge normalized to -1 or with two such 
fields $\theta,\bar{\theta}$ with charges -1 and 1 respectively. 
It is normally assumed that only 
the third generation of fermions have renormalizable couplings invariant
under U(1). The rest of the couplings arise in the effective theory from 
higher dimensional terms \cite{FN}:

$$ \Psi_i \Psi_j H ({\theta\over M})^{n_{ij}}$$
where $\Psi_i$ is a super-field, H is the Higgs doublet and M is some
higher mass scale which could be Planck scale $M_p$ and $n_{ij} =
\psi_i + \psi_j$ are positive numbers representing the charges of
$\Psi_i$, $\Psi_j$ under U(1) respectively. This gives rise to an $ij^{th}$ entry
of order $\lambda^{n_{ij}}$ in the mass matrix for the field $\Psi$. Identification
$\lambda\sim 0.22$ and proper choice of the U(1) charges leads to successful quark
mass matrices \cite{ramond,dudas1} this way.

{\it A priori} the model has eighteen free parameters: Fifteen U(1)
charges for quark and lepton super-fields, two U(1) charges for two
Higgs super-fields and the parameter $x = q_3 + d_3 + h_1$, which determines
the tan $\beta$ as, $ \tan \beta  \sim \l ^{x} (m_{t}/m_{b})$.  However, the requirements of 
reproducing  correct quark and lepton mass matrices 
along with the CKM matrix and the cancellations of the extra $U(1)$ anomalies 
(through the Green-Schwarz mechanism \cite{GS}) would reduce the number of
parameters to four. The appropriate values for the rest of the parameters
which are now written in terms of charge differences of form 
$q_{i3} = q_i - q_3$ etc, have been studied in the literature \cite{dudas1,chun}
and we present them here in Table {\rm I}
\newpage
\begin{center}
{\large Models}\\[20pt]
\end{center}
\begin{tabular}{|c|c|c|c|c|c|c|c|c|} \hline 
{\bf Models} & ${\bf l_{13}+e_{13}}$ & ${\bf l_{23}+e_{23}}$ & 
${\bf q_{13}}$ & ${\bf q_{23}}$ & $\bf u_{13}$ & $\bf u_{23}$ & 
${\bf d_{13}}$ & ${\bf d_{23}}$ \\ \hline \hline
{\rm IA} & 4 & 2 & 3 & 2  & 5 & 2 & 1 & 0 \\  \hline
{\rm IIA} & 4 & 2 & 4 & 3 & 4 & 1 & 1 & -1 \\ \hline
{\rm IIIA} & 4 & 2 & 4 & 2 & 5 & 2 & -1 & -1 \\ \hline
{\rm IVA} & 4 & 2 & -2 & -3 & -10 & 7 & 6 & 5 \\ \hline\hline
{\rm IB} & 5 & 2 & 4 & 3 & 4 & 1 & 1 & -1 \\ \hline
{\rm IIB} & 5 & 2 & 3 & 2 & 5 & 2 & 1 & 0 \\ \hline 
{\rm IIIB} & 5 & 2 & 4 & 3 & 4 & 1 & 1 & -1 \\ \hline
{\rm IVB} & 5 & 2 & -2 & -3 & -10 & 7 & 6 & 5 \\ \hline
\end{tabular}

\vskip 0.3 cm 

{\bf Table 1}: {\it In the above, Models in the class A reproduce exactly 
the lepton mass matrix, whereas models of class B take into 
consideration $O(\l)$ variations as the
$U(1)$ symmetry predictions are exact only upto $O(1)$.} 
\vspace{0.5cm}
Inclusion of the R-violating parameters can lead to additional constraints. 
Let us consider, as a starting point only 
lepton number violating couplings which are given as:
\beq
\label{super}
W_{{\not R}_p}= \l^{'}_{ijk}L_{i}Q_{j}D^{c}_{k} + \l_{ijk}
L_{i}L_{j}E^{c}_{k} + \e_{i} L_{i}H_{2}
\eeq
where we have used the standard notation. We will comment on the 
baryon number violating couplings later on. Under the $U(1)$ symmetry
the order of magnitudes of these couplings are also predicted, which
are constrained by low energy phenomenology. The most stringent constraint
comes from the possible choice of charges for the parameter $\e_i$. Similar
in nature to the $\m$ parameter, the $\e_i$ would have the following structure
under the $U(1)$ symmetry(if $l_i + h_2 \ge 0$):

\beq
\e_i \sim M \l^{l_i + h_2}
\eeq 
where $\l$ is the Cabibo angle. Unless the charges $l_i + h_2$ are appropriately
chosen, the predicted value for the $\e_i$ can grossly conflict with 
(a) The scale of $SU(2) \times U(1)$ breaking and 
(b) neutrino masses, as the
neutrino mass generated in the presence of these couplings 
is directly proportional
to $\e_i$. In order to prevent very large $\e_i$ one must ensure,\\
 a) $l_i + h_2 \ger 24$ 
(which has been derived by choosing $\l = 0.22$ and $M = 10^{16} GeV$ ) or\\
 b) $l_i + h_2 < 0$ so that the analyticity of $W$ will not allow such a term. 
This constraint has been so far neglected in the literature while considering
the structure of R-violation under the influence of the $U(1)$ symmetry. As we
will discuss below, this constraint plays an important role in deciding the
allowed patterns of R-violation. 

\section{ STRUCTURES OF TRILINEAR COUPLINGS}

In this section, we will consider the effect of the $U(1)$ symmetry on
the L-violating couplings, $\l'_{ijk}$, $\l_{ijk}$. The magnitudes and 
the structures of these couplings are determined by the following
equations:
 
\beqa
\label{lamb}
\lambda'_{ijk}&=& \theta(n'_{ijk})\lambda^{n'_{ijk}} \\
\lambda_{ijk}& =& \theta(n_{ijk})\lambda^{n_{ijk}} 
\eeqa
where $n'_{ijk} = c_i + n^{d}_{jk}$ , $n_{ijk} = c_i + n^{l}_{jk}$ ,
 $c_i=l_i+x+h_2-h$, $n^{d}_{jk}=q_{j3}+d_{k3}$, $n^{l}_{jk} = l_{j3} + e_{j3}$
with $n^{d}_{jk}$ and $n^{l}_{jk}$ being completely fixed for a given
model displayed in Table {\rm I}. The parameter $h = h_1 + h_2$. 
The analyticity requirement of the $W$ 
would lead to some of the trilinear couplings being zero if the corresponding
exponent is negative. As mentioned above, the low energy phenomenology has
stringent constraints on the magnitudes of these couplings \cite{allanach}. 
However, it turns out that the constraint from the $K^0 -\bar{K}^0$ mass
difference alone is sufficient to rule out the presence of the trilinear
couplings in most models. Here, we use a conservative estimate of the above
limit given as:
\beq
\label{koko}
\l'_{i12}\l'_{i21}\leq \l^{12}\sim 1.3 \cdot 10^{-8} 
\eeq

In addition to the above constraint one has to consider the constraints 
on the $\e_i$ parameters mentioned above, which leads to conditions (a) or (b). 
After taking these constraints into consideration,
the allowed structures of the trilinear couplings can be studied in the 
models presented in Table {\rm I}. 
Firstly, we observe from eq.(\ref{lamb}) that imposing the constraint
(a) $l_i + h_2 \ger 24$ would lead to trilinear couplings having orders
of magnitude $\sim \l^{19} \sim 10^{-12}$. Such a  small value of the
trilinear couplings would not have any phenomenological consequence. We will 
now consider imposing the second choice (b) $l_i + h_2 \leq 0$. 
To present the analysis in this case, we choose the phenomenologically 
most preferred model, namely model {\rm IA}. In this case, the trilinear
$\l'_{ijk}$ couplings are explicitly given as, 

\beq
 \l'_{ijk}  = \l^{l_i + h_2 + x}\left (
   \begin{array}{ccc}
       \l^{4} & \l^{3} & \l^{3} \\
       \l^{3} & \l^{2} & \l^{2} \\
       \l  &  1 & 1 
     \end{array}
     \right ) .
\eeq

\noindent where it is implicit that some element is zero if the corresponding 
exponent is negative. The above matrix is expressed inm terms of  the corresponding
matrix for the down-quarks, $\e (M_d)_{jk}$.  Hence,  for negative
$l_i+h_2$, it follows that the $\l'_{ijk}$ is either larger than the
matrix element $(M_d)_{jk}$ or is zero for every $i$. In the former case, 
the phenomenological requirement of the constraint, eq.(\ref{koko}) is not easily
met.  Specifically, equation for the $c_i$ gets translated to,
\beq
c_i \equiv l_i+h_2+x < -3\;\;\;\  {\mbox or} \geq \;\;\;3 
\eeq

This condition ensures that  $\l'_{i12}\l'_{i21}$ either
 satisfies  constraints from eq.(\ref{koko}) (when $c_i>3$ ) or is identically 
zero when $c_i<-3$. But $c_i\geq 3$ is untenable since $l_i+h_2\leq 0$ 
and $\tan\beta\sim \l^{x} (m_t/m_b) \geq O(1)$ needs
 $x\leq 2$ leading to $c_i\leq 2$. As a result one must restrict  $c_i$
to  less than -3 for {\it all} $i$.  The choice $c_i=-4$ is also ruled
out as mixing of fields in kinetic term would now produce the additional couplings which
are constrained by eq.(\ref{koko}).  Thus one concludes that only viable possibility 
from phenomenology is to require vanishing $\l'_{ijk}$ for all values of i,j,k.

The above arguments also serve to restrict the trilinear couplings $\l_{ijk}$.
Defining $(\Lambda_{k})_{ij} \equiv \l_{ijk}$,  we have, 
\beq
 (\Lambda_{1})_{ij}  = \l^{4} \left (
   \begin{array}{ccc}
       0 & \l^{c_2} & \l^{c_3} \\
       - \l^{c_2} & 0  & \l^{c_3 + l_2 - l_1} \\
       -\l^{c_3}  &  -\l^{c_3 + l_2 - l_1}  & 0
     \end{array}
     \right ),
\eeq

\beq
 (\Lambda_{2})_{ij}  = \l^{2} \left (
   \begin{array}{ccc}
       0 & \l^{c_1} & \l^{c_3 + l_1 - l_2} \\
       - \l^{c_1} & 0  & \l^{c_3} \\
       -\l^{c_3 + l_1 - l_2}  &  -\l^{c_3}  & 0
     \end{array}
     \right ),
\eeq

\beq
 (\Lambda_{3})_{ij}  = \left (
   \begin{array}{ccc}
       0 & \l^{c_2 + l_1 - l_3} & \l^{c_1} \\
       - \l^{c_2 + l_1 - l_3} & 0  & \l^{c_2} \\
       -\l^{c_1}  &  -\l^{c_2}  & 0
     \end{array}
     \right ),
\eeq

\noindent where $c_i$ are the same coefficients defined in the context of the $\lambda'$
and are required to be $<-4$ as argued above. It then immediately follows 
from the charge assignments of the Model {\rm IA}
that all the $\l_{ijk}$ except $\l_{123},\l_{231}$ and $\l_{312}$ are forced to 
be zero. Moreover, $\l_{312}$ and $\l_{231}$ cannot simultaneously be zero.

Thus, we have demonstrated an important conclusion that Model {\rm IA} can be 
consistent with phenomenology only if all $\lambda'_{ijk}$ and 
all $\lambda_{ijk}$ except at most two are zero. Similar exercise has been
done with the other models leading to similar conclusion. Taking into 
consideration all the constraints on the models, we have numerically looked
for integer solutions while restricting the absolute values of the charges
$q_3,~u_3,~d_3,~l_1,~l_2,~l_3$ to be $\le$10. This requirement is imposed for
simplicity. As an example, we present here the allowed values for the 
Model {\rm IIB} in the Table {\rm II}.

From the above analyses we can derive the following conclusions:

\noindent (1). While all $\l'_{ijk}$ are forced to be zero, 
 some models do allow one or two non-zero $\l_{ijk}$ which need not 
always be compatible with phenomenology.\\

\noindent (2) Although the term $L_iH_2$ is not directly allowed, 
it can be generated through the mechanism proposed 
by Giudice and Masiero \cite{GM}.  The order of magnitudes
 of the $\e_i$  given in this case can still be of phenomenological 
relevance \cite{rishi}.\\

\noindent (3)Though we have not considered the baryon number violating couplings, 
we have also not imposed baryon parity. Solutions show that the operator 
$U^{c}_{i} D^{c}_{j} D^{c}_{k}$ carries large -ve charge in all models. 
Thus Baryon number violation is automatically forbidden from
super-potential. It can also be shown that  the $\l''$ generated 
from effective U(1) violating D-term are extremely suppressed, O($10^{-15}$)
\cite{rishi}. So Proton stability is automatically explained in all models.\\

\begin{center}
%{\large Model IIB}\\[20pt]
\begin{tabular}{|c|c|c|c|c|c|c|c|c|c|c|c|} \hline
{\bf No.} & {\bf x} & {\bf q} & {\bf u} & {\bf d} & {$\bf l_1$} & {$\bf l_2$} & {$\bf l_3$}
& {$\bf f_1$} & {$\bf f_2$} & {$\bf f_3$} & {\bf If $\bf \lambda_{ijk}$
allowed} \\
\hline\hline

1 & 0 & 2 & 2 & -6 & -3 & -8 & -9 & -7 & -12 & -13 & No \\ 
2 & 0 & 2 & 3 & -7 & -8 & -5 & -7 & -13 & -10 & -12 & No \\ 
3 & 0 & 2 & 3 & -7 & -6 & -10 & -4 & -11 & -15 & -9 & No \\ 
4 & 1 & 2 & 3 & -6 & -8 & -2 & -7 & -13 & -7 & -12 & $\lambda_{231} \sim 
1.0 $\\ 
5 & 1 & 2 & 3 & -6 & -6 & -7 & -4 & -11 & -12 & -9 & No \\ 
6 & 2 & 2 & 3 & -5 & -6 & -4 & -4 & -11 & -9 & -9 & $\lambda_{231} \sim  
1.0$ \\
7 & 1 & 3 & 4 & -9 & -9 & -10 & -7 & -16 & -17 & -14 & No \\ 
8 & 2 & 3 & 4 & -8 & -9 & -7 & -7 & -16 & -14 & -14 & No \\ \hline

\end{tabular}
\end{center}
\vspace{0.3cm}
Table {\rm II}: Here we display the possible U(1) charge assignments
for the Model {\rm IIB} of Table I, consistent with the
phenomenological constraints listed in the text. Absolute values of
$q_3,u_3,d_3,l_1,l_2,l_3$ have been restricted to $\le$ 10 for simplicity.

\section{SUMMARY}
The supersymmetric standard model allows 39 lepton number violating parameters
which are not constrained theoretically. We have shown in this talk that the 
U(1) symmetry, invoked to understand fermion masses can play an important role 
in constraining these parameters. 
%We restricted ourselves to integer U(1) charges and 
%considered  many different U(1) charge assignments compatible with fermion spectrum.
We have shown that  only phenomenologically consistent possibility, in this context 
is that all the trilinear $\l'$ and all but two $\l$ couplings to be zero or
extremely small of $O(10^{-4})$. While the patterns of $R$ violation have been earlier 
discussed in the presence of U(1) symmetry, the systematic confrontation of these
pattern with phenomenology leading to this important conclusion was not made to 
the best of our knowledge.This way, U(1) symmetry is shown to require that only
four or five of the total 39 lepton number violating couplings could have magnitudes
in the phenomenologically interesting range! 

\vspace{0.8cm}
{\large {\bf Acknowledgments}}

\vspace{0.3cm}
I thank the organizers of $XIV$ DAE Symposium on High Energy Physics, for
giving me an opportunity to present this work.


\begin{thebibliography}{99}
\bibitem{rneutrino} L. Hall and M. Suzuki \np{B231}{419}{84};
A. S. Joshipura and M. Novakowski \pr{D51}{2421}{95}; 
A. S. Joshipura and S. K. Vempati \pr{D60}{095009}{99}; {\it ibid}
\pr{D60}{111303}{99};
A. S. Joshipura, R. D. Vaidya and S. K. Vempati
\prd{D65}{053018}{02} [arXiv: he-ph/0107204]; {\it ibid} to appear in
Nuclear Physics B [arXiv: hep-ph/0203182]. 
\bibitem{FN}C. D. Frogatt and H. B. Neilsen,  \np {B147}{79}{277}
\bibitem{rishi} R. D. Vaidya, A. S. Joshipura and S. K. Vempati, \prd {D 62}{00}{093020}
\bibitem{ramond} P. Binetruy and P. Ramond, \pl {B 350}{95}{49}
\bibitem{dudas1} E. Dudas, S. Pokorski and C. A. Savoy, \pl {B 356}{95}{45}.
\bibitem{GS} M. B. Green and J. H. Schwarz, \pl {B149}{84}{117},
\bibitem{chun} E. J. Chun and A. Lukas, \pl {B387}{96}{99}, E. J. Chun,
K. Choi and H. Kim \pl{B394}{89}{97} [arXiv:hep-ph/9611293].
\bibitem{allanach} For a recent review see, B. Allanach, A. Dedes and H. Driener, 
\pr {D60}{99}{075014}.
\bibitem{GM} G. F. Giudice and A. Masiero, \pl {B206}{88}{480}.


\end{thebibliography}
\end{document}